\documentclass[11pt]{article}
\usepackage{abstract}
\usepackage[a4paper,margin=1in]{geometry}
\usepackage{amsmath,amssymb}
\usepackage{microtype}
\usepackage{enumitem}
\usepackage{hyperref}
\usepackage{booktabs} 
\usepackage{graphicx}
\hypersetup{
  colorlinks=true, linkcolor=blue, citecolor=blue, urlcolor=blue,
  pdftitle    ={An intersectional informational data valuation theory},
  pdfkeywords ={Data pricing, Mutual information, Externalities, Intersectionality, Privacy economics}
}

\newcommand{\I}{\mathrm{I}}

\begin{document}
\title{\textbf{An Information-Theoretic Intersectional Data Valuation Theory}}
\author{Eduardo C. Garrido-Merchán\\ 
    Universidad Pontificia Comillas \\
    Instituto de Investigación Tecnológica \\
    \texttt{ecgarrido@comillas.edu}}        
\date{July 2025}
\maketitle

\begin{abstract}
In contemporary digital markets, personal data often reveals not just isolated traits, but complex, intersectional identities based on combinations of race, gender, disability, and other protected characteristics. This exposure generates a privacy externality: firms benefit economically from profiling, prediction, and personalization, while users face hidden costs in the form of social risk and discrimination. We introduce a formal pricing rule that quantifies and internalizes this intersectional privacy loss using mutual information, assigning monetary value to the entropy reduction induced by each datum. The result is a Pigouvian-style surcharge that discourages harmful data trades and rewards transparency. Our formulation has the advantage that it operates independently of the underlying statistical model of the intersectional variables, be it parametric, nonparametric, or learned, and can be approximated in practice by discretizing the intersectional joint probability distributions. We illustrate how regulators can calibrate this surcharge to reflect different societal values, and argue that it provides not just a technical fix to market failures, but also a redistributive shield that empowers vulnerable groups in the face of asymmetric digital power.
\end{abstract}
\begin{quote}
\emph{“Only by being a man or woman for others does one become fully human.”} \\
\hfill — Pedro Arrupe, SJ.
\end{quote}
\section{Introduction}
Digital platforms monetise behavioural traces, yet the advertised market price rarely reflects the \emph{social cost} of revealing protected attributes.  
For example, knowing that a user opens a fitness app at 6:03~a.m.\ every weekday appears innocuous, until one realises it sharply increases the probability that the user is a young able‑bodied male, intersecting age, disability and gender.  
Traditional pricing schemes ignore that externality; consequently, data markets over‑produce privacy harm.

This imbalance is not merely a technical oversight but a structural economic failure.  
In contemporary data capitalism \cite{Zuboff2019}, platforms extract surplus value not from labour, but from human behavioural data, especially from individuals with little bargaining power or digital literacy.  
Concretely, we can interpret data subjects, particularly those belonging to historically oppressed or racialised groups, as being dispossessed of the informational value that their digital traces produce.  
The expropriation of this data capital happens quietly, through default app permissions, dark patterns, and a lack of meaningful opt-out mechanisms \cite{Eubanks2018}.  
While users are offered free services, platforms profit from opaque inference pipelines that reconstruct intimate protected traits from trivial inputs: location pings, typing speed, or even silence.

Intersectionality \cite{Crenshaw1989} insists that harm compounds across overlapping identities.  
This has material implications in data markets.  
Empirical studies show that automated inference systems disproportionately harm women of colour, the disabled poor, and other multiply marginalised subjects \cite{Noble2018,Barocas2019}.  
Yet their data is often the cheapest to collect, precisely because platforms deploy such systems in low-consent or high-vulnerability contexts: public assistance portals, gig economy apps, or default Android settings in the Global South.  
As a result, marginalised individuals not only receive fewer benefits from the data economy but bear its heaviest algorithmic burdens.

From a competition policy perspective, the unchecked concentration of data power in Big Tech platforms creates conditions ripe for cartelisation.  
Data monopolies benefit from economies of scale and scope, allowing them to infer protected attributes with ever-finer granularity and ever-lower marginal cost per user.  
This drives a wedge between private gains and social harms, violating basic tenets of welfare economics \cite{Stiglitz2000}.  
In particular, Pigouvian theory suggests that markets with externalities require corrective pricing: harmful outputs, such as privacy violations, must be taxed or compensated at a rate proportional to their marginal social cost.

We take the standpoint of a regulator or platform that wants an implementable, quantitative rule:

\begin{quote}
\textit{How much compensation or Pigouvian surcharge should accompany the sale of a single datum, given its ability to reveal the protected intersectional profile of the data subject?}
\end{quote}

Our answer hinges on a single information‑theoretic observable: the expected drop in Shannon entropy of the protected profile $S$ once the datum $X$ is known.  
Because that drop equals the mutual information $\I(X;S)$, the pricing rule is readily measurable with modern predictive models.  
This transforms privacy from a qualitative norm into an economically priced quantity, grounded in probabilistic inference and social justice.

The remainder of this paper is organised as follows.  
Section~\ref{sec:related_work} reviews related work at the intersection of data economics, privacy valuation, and intersectional fairness, identifying key gaps in existing pricing models and ethical frameworks. After that, Section 3 introduces the fundamentals of the intersectional theory. Section 4 introduces a set of five applied axioms that formalise the economic and social requirements for a defensible data pricing scheme under intersectional risk. Section 5 derives the unique valuation rule consistent with these axioms, showing that it must be proportional to the mutual information between the datum and the protected profile. Section 6 interprets the resulting pricing rule in terms of Pigouvian economics, informational asymmetry, and power concentration in digital markets. Section 7 offers practical guidance on estimating the proposed measure using predictive models and discusses how regulatory bodies might calibrate the price to reflect normative priorities. We conclude in Section 8 with reflections on the ethical implications of informational capitalism and the potential of intersectional data valuation to reorient data markets toward justice.

\section{Related Work}\label{sec:related_work}

The problem of data valuation has gained traction across several disciplines, from computer science to law and economics.  
In the domain of privacy economics, early contributions focused on individuals’ willingness to pay to protect their data or willingness to accept compensation for its disclosure \cite{Acquisti2005}.  
More recent approaches have framed privacy as an externality requiring Pigouvian correction, particularly in the context of digital platforms with informational asymmetries \cite{Jones2020}.  
However, most models assume homogeneous privacy preferences and fail to account for structural disparities tied to race, gender, class or disability.

The growing literature on data markets has proposed pricing rules based on marginal predictive value \cite{Ghorbani2020}, Shapley value approximations \cite{Jia2019}, or consumer surplus under specific demand models \cite{Bergemann2015}.  
These frameworks are rarely sensitive to \emph{who} is represented in the data, and often treat all users as interchangeable agents in utility maximisation schemes.

From a fairness perspective, recent works have proposed data deletion or perturbation methods that minimise disparate impact \cite{Feldman2015}, or that use differential privacy to guarantee statistical indistinguishability \cite{Dwork2014}.  
Still, they do not translate these concepts into pricing mechanisms that could shape market incentives.

Finally, few works explicitly integrate intersectionality into the valuation process.  
Exceptions like \cite{Kattan2022} explore the relevance of protected attributes in fairness-aware data valuation, but stop short of formulating an axiomatic economic theory for pricing data under intersectional risk.

Our work fills this gap by grounding data valuation in information theory while explicitly tying price to the expected disclosure of a protected intersectional profile, thus connecting technical measurement, economic justice, and social theory.

\section{Intersectional Privacy: A Theoretical Foundation}

The theory of intersectionality, originally formulated by Crenshaw~\cite{Crenshaw1989}, posits that individual experiences of oppression and vulnerability are not adequately captured by analyzing isolated demographic categories (such as sex, race, or disability) in isolation. Rather, these experiences are shaped by the interaction of multiple identity dimensions, each of which compounds and modulates the impact of the others. In data contexts, this implies that privacy risks are not additive across protected attributes, but emerge from their \emph{joint distribution}.

Classical privacy metrics often fail to account for these structural dependencies. For example, measuring the information leakage about gender or ethnicity individually may underestimate the exposure experienced by individuals situated at vulnerable intersections—such as women with disabilities or ethnic minorities with specific health conditions. Hence, privacy assessments must be extended from marginal attributes to entire vectors of protected characteristics, modeled as multivariate random variables $S = (S_1, S_2, \ldots, S_k)$.

From an information-theoretic perspective, the mutual information $\I(X; S)$ provides a natural and general measure of how much a public observable $X$ reveals about the full protected profile $S$. Critically, $\I(X; S)$ captures dependencies across all combinations of values of $S$, regardless of whether the components are categorical, ordinal, or continuous. This formulation respects the intersectional insight that vulnerability is often linked to specific value combinations, for example being female, immigrant and disabled rather than to each attribute in isolation. We illustrate in Figure 1 the variables included in the intersectional paradigm. 

\begin{figure}[h]
  \centering
  \includegraphics[width=0.99\linewidth]{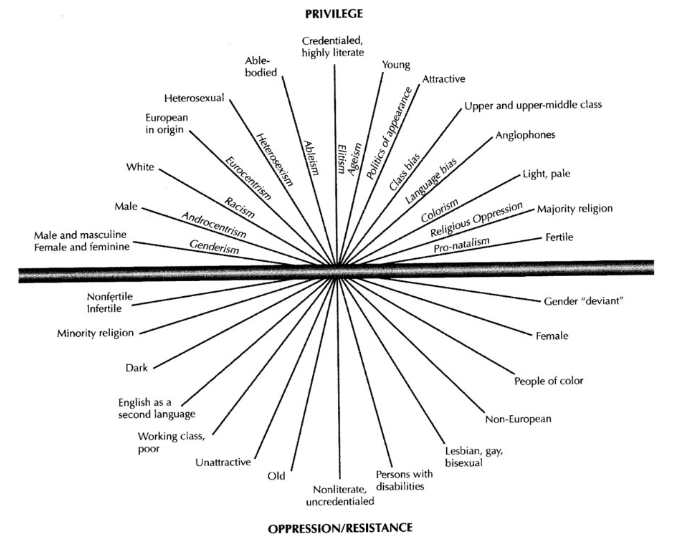}
  \caption{Variables included in the intersectional paradigm. We emphasize that it is important not only to consider the marginal variables but their joint distributions.}
  \label{fig:autism_knowledge}
\end{figure}

Regulatory frameworks aiming to align data practices with intersectional justice must therefore require both the monitoring and minimization of joint information leakage. In particular, privacy impact assessments and algorithmic audits should quantify not only the leakage of individual sensitive attributes, but also the inferential risk associated with their conjunctions. Failing to do so risks reinforcing structural inequalities through seemingly neutral technical systems.

In this work, we adopt mutual information as a quantitative tool to integrate intersectionality into economic and regulatory models of privacy. This allows us to assign monetary cost to intersectional exposure and to internalize it through Pigouvian surcharges or compensation mechanisms, thus bridging formal economic reasoning with anti-discriminatory ethical imperatives.

\section{Axiomatization of the Theory}\label{sec:axioms}
This section establishes the formal foundations of our pricing mechanism through a minimal and interpretable set of axioms. Rather than anchoring the model in any specific economic or legal theory, we adopt a general information-theoretic lens that quantifies social harm via reductions in uncertainty about protected attributes. The axioms presented are empirically verifiable, model-agnostic, and fully compatible with existing data practices. They reflect intersectional concerns by making mutual information $\I(X;S)$ the sole channel through which observables affect social cost, thereby ensuring that privacy externalities are priced independently of market-specific utility. This axiomatization serves as the normative backbone of the entire framework.

We first begin this section describing the notation that is going to be used for clarity purposes $S=(S_1,\dots,S_m)$ is the discrete vector of protected attributes (race, gender, disability, …). $X$ is the datum offered for sale (timestamp, GPS ping, purchase record, \emph{etc.}) $H(S)$ is the Shannon entropy of $S$ under the analyst’s \emph{prior} belief $P_S$. $H(S|X)$ is the posterior entropy after observing $X$. $\I(X;S)=H(S)-H(S|X)$ is the mutual information, i.e.\ expected entropy drop. $c_p$ is the marginal \emph{production} cost of collecting/storing $X$ (electricity, servers, labour). $\lambda$ is the societal cost per nat\footnote{One nat = $\ln 2$ bits.  Any other monotone transformation of information units would simply rescale $\lambda$ as we will later explain.} of uncertainty reduction. We illustrate in Figure 2 how a single datum can affect the conditional joint probability distribution of intersectional variables having its entropy dropped.

\begin{figure}[h]
  \centering
  \includegraphics[width=0.99\linewidth]{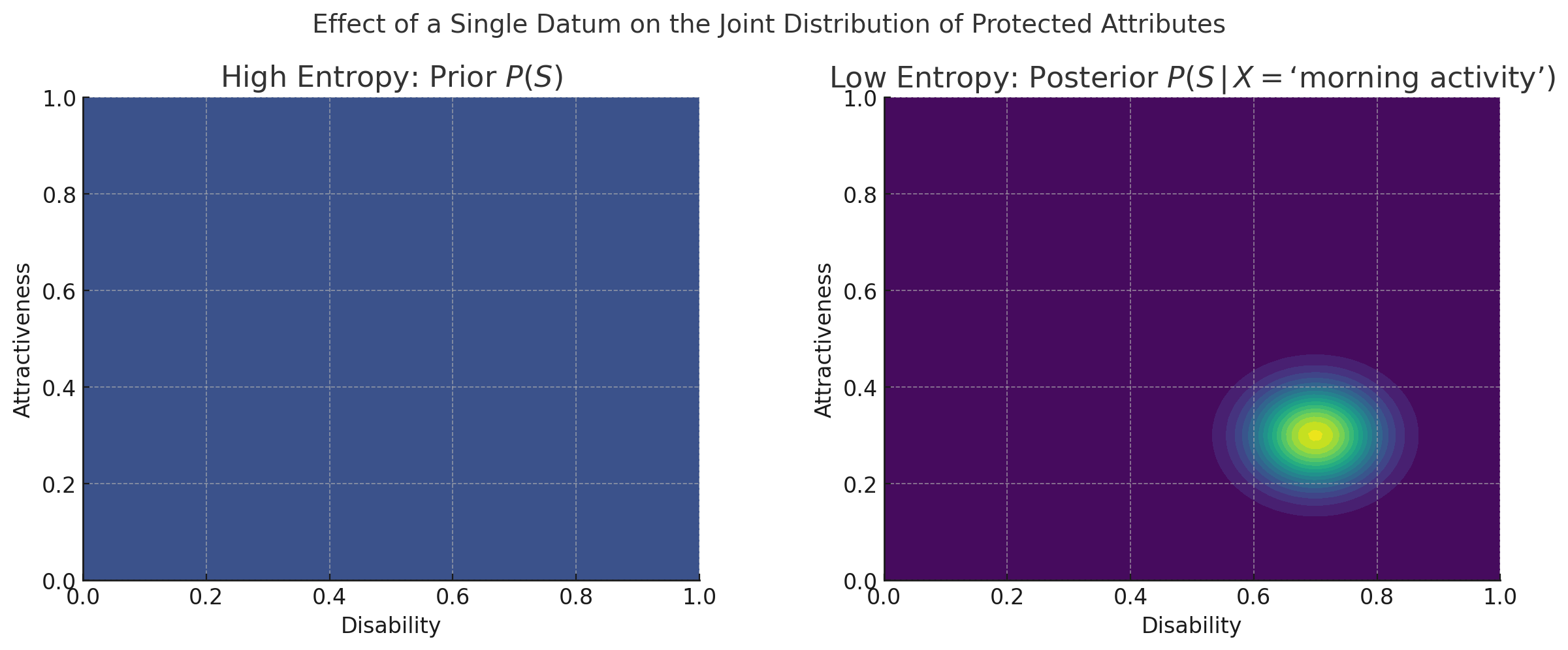}
  \caption{Reduction in entropy on a joint probability distribution belonging to the set of intersectional variables that produces a single marginal datum. We can see that the marginal joint distribution on the left has an uniform prior that represents complete uncertainty and maximum entropy of the pair of variables. Having observed the morning activity of an user, the conditional distribution of its disability and attractiveness changes, having more knowledge and less entropy once the datum is known, showing the privacy of the user to the company.}
  \label{fig:autism_knowledge}
\end{figure}

Let $V(X)$ denote the socially fair \emph{price tag} attached to one observation of $X$.  
That price may be paid to the user, to a compensation fund, or to the treasury (Pigouvian tax).  
We state axioms every practitioner can empirically verify.

\begin{enumerate}[label=\textbf{A\arabic*},leftmargin=*]

\item\textbf{(Intersectional Externality)}  
  The only social harm considered is the uncertainty reduction about the entire protected profile $S$.  
  Therefore \(V\) can depend on $X$ \emph{only} through $\I(X;S)$.

\item\textbf{(No‑Harm, No‑Cost)}  
  If $X$ is independent of $S$ \emph{in the analyst’s predictive model}, then $\I(X;S)=0$ and \(V(X)=c_p\).  
  The user pays (or is paid) nothing beyond the production cost.

\item\textbf{(Marginal Information Monotonicity)}  
  If two candidate data items $X$ and $Y$ satisfy $\I(X;S)>\I(Y;S)$ under the same model, then
  \(
      V(X) > V(Y).
  \)

\item\textbf{(Additivity for Independent Observations)}  
  For independent observations $X_1,X_2$ collected sequentially,
  \[
    V(X_1,X_2) \;=\; V(X_1) \;+\; V(X_2).
  \]
  This ensures pricing scales linearly with the number of independent disclosures.

\end{enumerate}

Together, these axioms uniquely characterize a simple yet powerful pricing rule grounded in information theory: social cost scales linearly with the expected reduction in uncertainty about intersectional identity. The resulting valuation function $V(X)$ is fully general, interpretable, and robust to changes in modeling assumptions. By axiomatizing privacy as an informational externality and tying compensation directly to $\I(X;S)$, we bridge the gap between fairness principles, regulatory implementation, and market compatibility. This foundational structure ensures that the rest of the framework inherits both conceptual clarity and operational feasibility.

\section{Information theoretic Valuation of Intersectionality Theorem}
We now formalize the consequences of the axioms previously stated. Following a standard approach in information theory and functional equations, we show that these axioms uniquely determine a linear pricing rule. The result below captures the core insight of the framework: that the social value of a datum must increase proportionally to the amount of protected information it reveals, measured via mutual information.

\paragraph{Theorem.}
\emph{Under Axioms A1–A4, following information's theory proofs, the unique\footnote{Up to currency scaling by a positive constant.} pricing rule compatible with them is}
\begin{equation}\label{eq:price_rule}
    V(X) \;=\; c_p \;+\; \lambda \, \I(X;S),
\end{equation}
where $V(X)$ is the economic value (or cost) of the observable $X$, $c_p$ is its internal marginal cost, $\I(X;S)$ is the mutual information between $X$ and the full vector of protected attributes $S$, and $\lambda > 0$ is a policy-determined Pigouvian multiplier expressing the marginal social cost per nat of leaked information. We illustrate in Figure 3 a representation of the linear social cost that would correspond to the pigouvian tax inferred by the formula for different values of $\lambda$.  

\begin{figure}[h]
  \centering
  \includegraphics[width=0.99\linewidth]{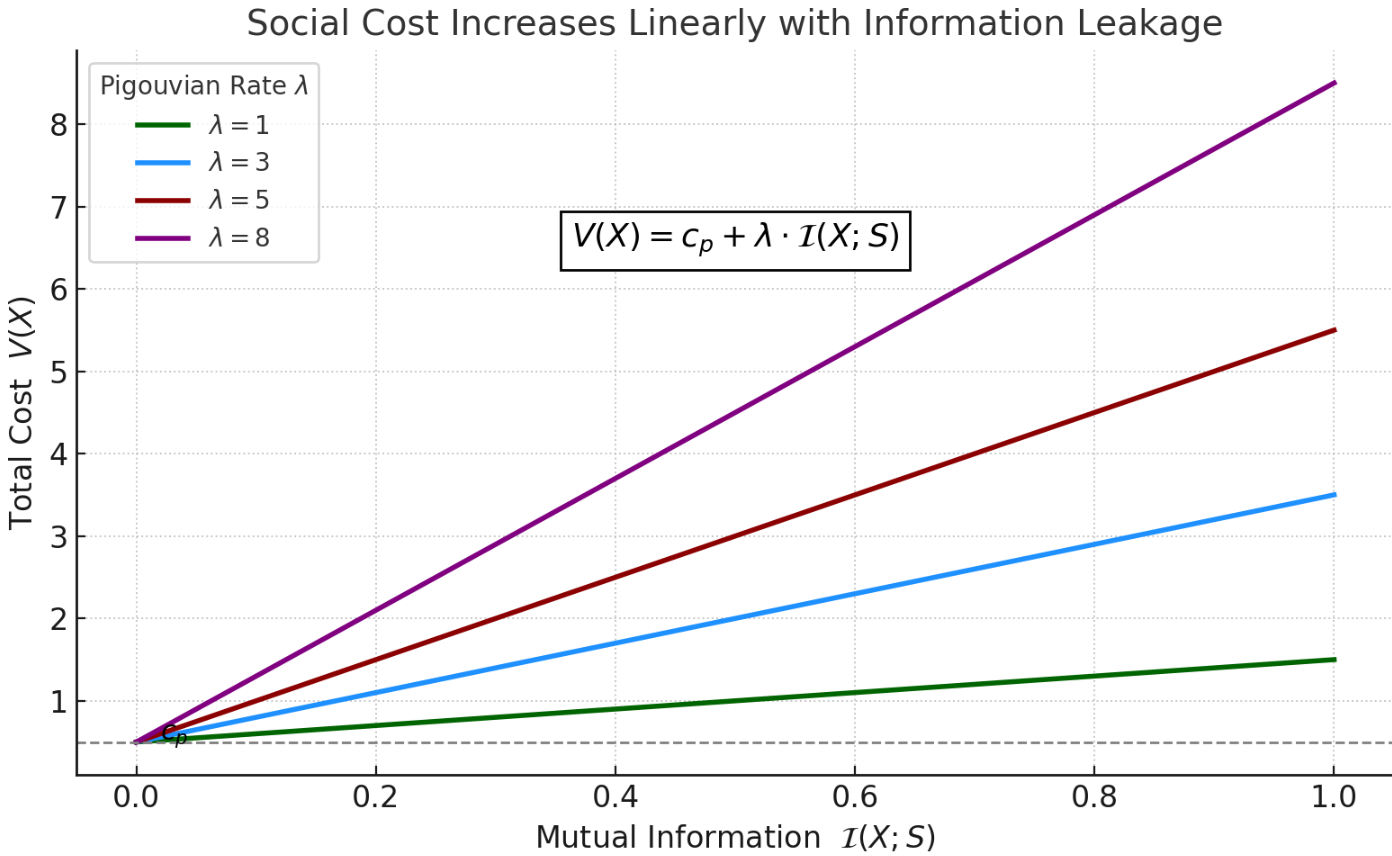}
  \caption{Illustration of the proposed pricing rule, where the total cost $V(X)$ increases linearly with the mutual information $\mathcal{I}(X;S)$ revealed by a datum. Different slopes $\lambda$ reflect varying regulatory valuations of privacy, while $c_p$ represents the baseline production cost.}
  \label{fig:autism_knowledge}
\end{figure}

\paragraph{Intuitive Proof}
Axiom A1 restricts the valuation function $V$ to depend only on the information leakage $\I(X;S)$, i.e., $V = f(\I)$.  
A2 imposes the normalization condition $f(0) = c_p$, ensuring that when $X$ leaks no protected information, its price equals its internal cost.  
A3 requires monotonicity: more information leakage must increase the cost, so $f$ is strictly increasing.  
A4 imposes additivity over independent observables, i.e., $f(\I_1 + \I_2) = f(\I_1) + f(\I_2) - c_p$.  
Hence, the only class of strictly increasing functions satisfying these conditions and following information theory is:  
\[
f(\I) = c_p + \lambda \I.
\]

\paragraph{Interpretation as Uniform Weighting of Intersectional Risk.}

The formulation in Equation~\eqref{eq:price_rule} treats the mutual information $\I(X;S)$ as a scalar summary of the disclosure risk posed by $X$ with respect to the protected profile $S$. However, in practice, $S$ comprises multiple protected variables—such as sex, ethnicity, disability, age, and religion—some of which may be continuous or jointly distributed. The mutual information $\I(X;S)$ captures leakage over the \emph{entire joint distribution} of $S$, including all marginals and higher-order intersections.

Importantly, when we compute $\I(X;S)$ in this setting, we are effectively averaging the reduction in entropy across all marginal and joint distributions of the components of $S$, assigning equal importance to each distinguishable configuration of $S$. This can be seen by rewriting the cost rule in a disaggregated form:
\[
    V(X) = c_p + \lambda \cdot \sum_{i=1}^m w_i \, \I(X;S_i),
\]
where $\{S_i\}$ includes all univariate marginals (e.g., $\text{sex}$), bivariate intersections (e.g., $(\text{sex},\text{disability})$), and so forth up to the full joint $S$, and where $w_i$ represents a normalized weighting over these combinations. The standard formulation in~\eqref{eq:price_rule} corresponds to the special case where all $w_i$ are equal—i.e., a uniform prior over all intersectional configurations—and the same $\lambda$ applies to all.

This implies that the current model encodes a form of \emph{egalitarian intersectional valuation}, in which all dimensions and their interactions are considered equally significant. However, this is not the only possible formulation. Future versions of the pricing rule could allow:
\[
    V(X) = c_p + \sum_{i=1}^m \lambda_i \, \I(X;S_i),
\]
with heterogeneous multipliers $\lambda_i$ to reflect regulatory priorities or empirically grounded risks (e.g., weighting disability-related disclosure more heavily in healthcare contexts, or ethnicity in political advertising). Such an extension would make the surcharge more context-sensitive and aligned with domain-specific fairness goals.

\paragraph{Units and Conversion.}

The proposed surcharge expresses informational risk in monetary units per nat, e.g., euros per nat or dollars per nat. If a regulator or system designer wishes to express the cost per bit instead of per nat, one may apply a logarithmic conversion constant:
\[
    \lambda_{\text{bit}} = \lambda_{\text{nat}} \cdot \log_e(2) \approx 0.693 \cdot \lambda_{\text{nat}}.
\]
Similarly, currency conversion can be introduced by multiplying by a fixed exchange rate $\rho$ (e.g., from USD to EUR). The combined conversion yields:
\[
    \lambda^{(\text{EUR/bit})} = \rho \cdot \log_e(2) \cdot \lambda^{(\text{USD/nat})}.
\]
Thus, although $\lambda$ is calibrated in a specific unit system, all variations across entropy bases and currency systems reduce to multiplicative constants, preserving the linearity and interpretability of the model.

\paragraph{Relative Exposure Ratio.}

In addition to pricing absolute information gain, one may define a normalized \emph{exposure ratio} as
\[
    r(X; S) = \frac{\I(X; S)}{H(S)},
\]
which quantifies the fraction of total entropy in the protected intersectional profile $S$ that is revealed by the observable $X$. This dimensionless ratio lies in $[0,1]$ and reflects the relative informativeness of $X$ with respect to the maximum possible disclosure. It can serve as the basis for an alternative pricing rule grounded in proportional risk, where the surcharge is expressed as a fraction of a fixed statutory maximum penalty $\Pi_{\text{max}}$:
\[
    V(X) = c_p + r(X; S) \cdot \Pi_{\text{max}}.
\]
This formulation maintains interpretability, respects upper legal bounds, and offers a normalized mechanism for taxation that adjusts across populations with different baseline entropy levels. We illustrate the pigouvian tax according to the value of $\Pi_{\text{max}}$ in Figure 4.

\begin{figure}[h]
  \centering
  \includegraphics[width=0.99\linewidth]{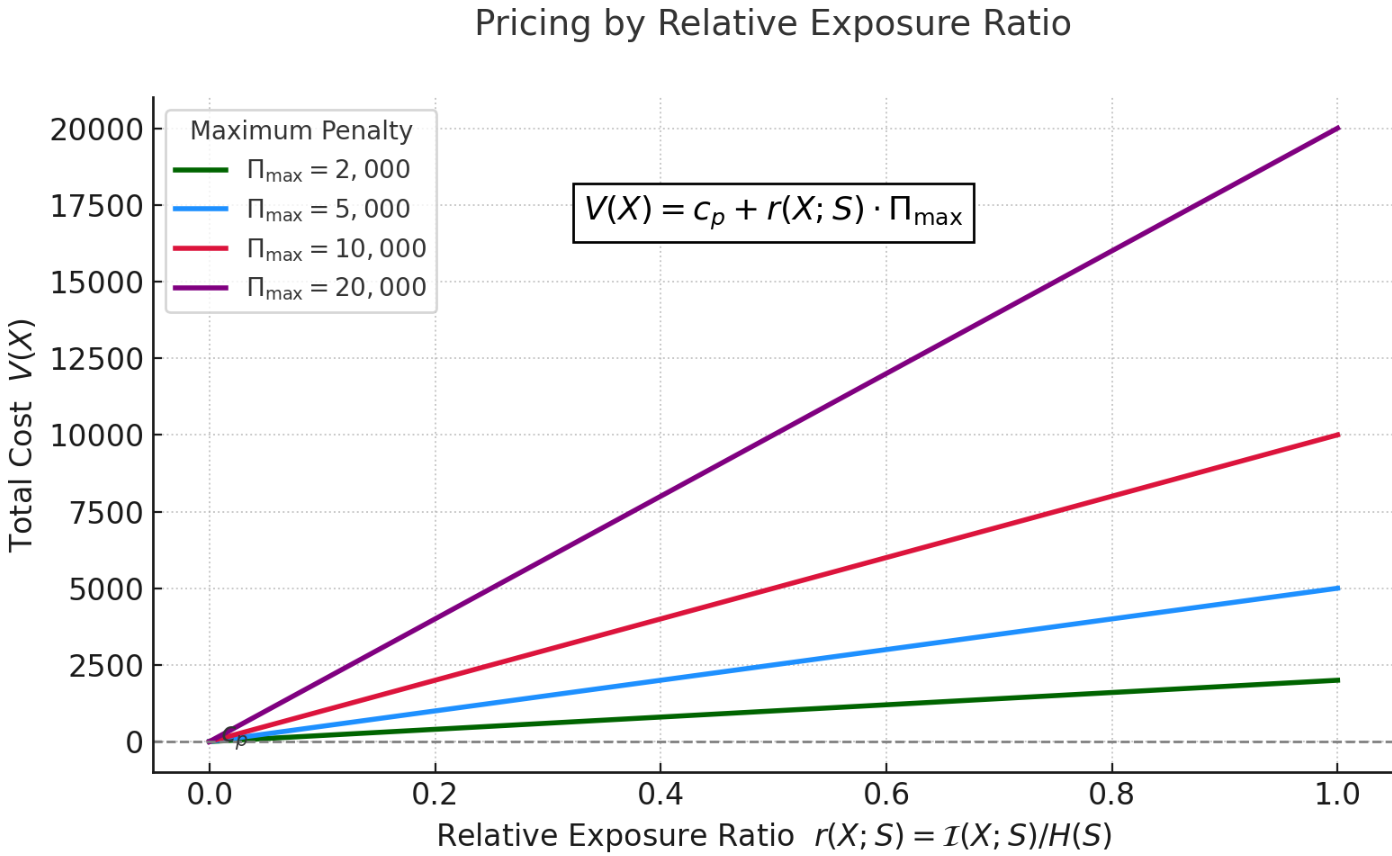}
  \caption{Relative Exposure Pricing Rule.  
Each line shows the total data value \(V(X)\) as a function of the relative exposure \(r(X;S)\), for different values of the statutory maximum penalty \(\Pi_{\text{max}}\).  
The formula \(V(X) = c_p + r(X;S) \cdot \Pi_{\text{max}}\) defines a normalized, interpretable surcharge rule that scales proportionally with privacy risk.}
  \label{fig:autism_knowledge}
\end{figure}

\paragraph{Model-Agnostic Applicability.}

A key strength of the proposed pricing mechanism is its independence from the specific form of the probabilistic model used to estimate the joint distribution $P(X, S)$ or the conditional $P(S \mid X)$. Whether the underlying model is parametric (e.g., a mixture of Gaussians), econometric (e.g., logistic or probit regression), nonparametric (e.g., kernel density estimators), based on dimensionality-reduced abstract variables (e.g., via PCA or autoencoders), or implemented via high-capacity deep learning architectures, all such models aim to approximate a distribution from which mutual information can be derived. Since the Pigouvian surcharge depends solely on the quantity $\I(X; S)$—regardless of how $P(X, S)$ is internally represented—the taxation scheme remains valid across modeling paradigms. This universality ensures that the policy is both statistically grounded and operationally robust, providing a consistent framework for regulating informational externalities in real-world systems. Even in Bayesian or implicitly generative settings—such as variational inference or generative adversarial networks—$\I(X;S)$ can be approximated via Monte Carlo sampling from learned distributions, further reinforcing the model-agnostic nature of the proposed mechanism.

\subsection*{Illustrative Example for a single variable}

Consider a simplified population where the protected attribute $S$ is binary: \textit{male} or \textit{female}.  
Suppose a datum $X$ represents the time of day (morning or evening) when a person opens a health app. Assume that this variable is independent of all the other variables belonging to the intersectional paradigm.
Let us assume the joint distribution $P_{X,S}$ is given by:

\begin{center}
\begin{tabular}{c|cc}
    & $S=$ male & $S=$ female \\
\hline
$X=$ morning & 0.30 & 0.10 \\
$X=$ evening & 0.20 & 0.40 \\
\end{tabular}
\end{center}

From this, we compute the marginal distributions:

\begin{align*}
    P(S=\text{male}) &= 0.30 + 0.20 = 0.50, \\
    P(S=\text{female}) &= 0.10 + 0.40 = 0.50, \\
    P(X=\text{morning}) &= 0.30 + 0.10 = 0.40, \\
    P(X=\text{evening}) &= 0.20 + 0.40 = 0.60.
\end{align*}

The prior entropy of $S$ is:

\[
H(S) = -\sum_{s} P(s) \log_2 P(s) = -2 \times 0.5 \log_2 0.5 = 1 \text{ bit}.
\]

The conditional entropy $H(S|X)$ is computed as:

\begin{align*}
H(S|X) &= P(x=\text{morning}) \cdot H(S|x=\text{morning}) + P(x=\text{evening}) \cdot H(S|x=\text{evening}) \\
       &= 0.4 \cdot H\left( \frac{0.30}{0.40}, \frac{0.10}{0.40} \right) + 0.6 \cdot H\left( \frac{0.20}{0.60}, \frac{0.40}{0.60} \right) \\
       &= 0.4 \cdot H(0.75, 0.25) + 0.6 \cdot H(0.\overline{3}, 0.\overline{6}) \\
       &= 0.4 \cdot \left[-0.75 \log_2 0.75 - 0.25 \log_2 0.25\right] \\
       &\quad + 0.6 \cdot \left[-(1/3) \log_2 (1/3) - (2/3) \log_2 (2/3)\right] \\
       &\approx 0.4 \cdot 0.811 + 0.6 \cdot 0.918 \\
       &\approx 0.864 \text{ bits}.
\end{align*}

Therefore, the mutual information is:

\[
I(X;S) = H(S) - H(S|X) \approx 1 - 0.864 = 0.136 \text{ bits}.
\]

Suppose we calibrate the externality price to $\lambda = \$10^5$ per bit of information obtained for this variable belonging to the set of intersectional variables.  
Then the intersectional value of the datum $X$ is:

\[
V(X) = c_p + \lambda \cdot I(X;S) = c_p + 10^5 \cdot 0.136 = c_p + \$13,600.
\]

This means that even a seemingly minor behavioural trace, like app usage time, may carry significant social cost due to its power to infer protected traits.  
A market respecting this valuation would either compensate the user accordingly, or disincentivise the use of such data unless strictly necessary.

\subsection*{Illustrative Example: Intersecting Protected Attributes}

To illustrate how intersectional attributes can be jointly revealed by a model, consider a protected variable $S = (\text{sex}, \text{disability})$ defined over four categories: (male, abled), (male, disabled), (female, abled) and (female, disabled). Let $X$ represent a publicly observable variable (e.g., time preference: morning or evening). The joint distribution $P(X, S)$ is given below:

\begin{center}
\begin{tabular}{c|cccc}
\toprule
$X$ & (male, abled) & (male, disabled) & (female, abled) & (female, disabled) \\
\midrule
morning & 0.25 & 0.05 & 0.05 & 0.05 \\
evening & 0.15 & 0.05 & 0.25 & 0.15 \\
\bottomrule
\end{tabular}
\end{center}

This example shows that a marginal analysis of either sex or disability alone would obscure key dependencies. For instance, observing $X = \text{morning}$ reveals a strong likelihood that the subject is male and abled (0.25 out of 0.40), whereas the probability of being female and abled is only 0.05 out of 0.40. Such asymmetric inferential power is only visible when the joint distribution is considered, and would be missed by measuring $\I(X; \text{sex})$ or $\I(X; \text{disability})$ separately.

Consequently, regulatory assessments must operate over the full joint entropy $H(S)$ and mutual information $\I(X; S)$, where $S$ encompasses all protected attributes in combination. This is especially crucial in intersectional contexts where vulnerability arises not from individual attributes but from their interaction.

\subsection*{Example: Revealing a Continuous Intersectional Attribute via a Trivial Observable}

Consider a case where the protected variable $S$ is a continuous-valued degree of motor impairment, ranging from $0$ (no disability) to $1$ (severe motor disability). As we will later see, for practical purposes, it is much easier to just discretize the continuous variables and create a discrete joint distribution of intersectional variables. However, the tax can also be estimated in the case of continuous variables with the following methodology. Let $X$ be a seemingly innocuous observable, such as average keystroke interval (in milliseconds), recorded during online form input.

Assume we collect a dataset of $n = 1000$ samples $(x_i, s_i) \sim P(X, S)$ from real-world usage. The protected attribute $S$ is not directly observable to the system designer but is correlated with $X$ due to slower typing speed among users with more severe motor impairments.

To estimate the privacy leakage $\I(X; S)$, we proceed in two steps:

\paragraph{Step 1: Estimate the joint and marginal distributions.}
We apply kernel density estimation (KDE) with Gaussian kernels to estimate $p(x,s)$, $p(x)$, and $p(s)$:
\[
    \hat{p}_h(x, s) = \frac{1}{n} \sum_{i=1}^{n} \mathcal{K}_h(x - x_i) \mathcal{K}_h(s - s_i),
\]
where $\mathcal{K}_h$ is a Gaussian kernel with bandwidth $h$ selected via cross-validation or Silverman's rule.

\paragraph{Step 2: Monte Carlo approximation of mutual information.}
We estimate the mutual information using the following identity:
\[
    \I(X; S) = \mathbb{E}_{(X,S)} \left[ \log \frac{p(X,S)}{p(X)p(S)} \right].
\]
This expectation is approximated by Monte Carlo integration:
\[
    \widehat{\I(X; S)} = \frac{1}{n} \sum_{i=1}^{n} \log \frac{\hat{p}_h(x_i, s_i)}{\hat{p}_h(x_i)\hat{p}_h(s_i)}.
\]

In a representative simulation using synthetic but realistic data, we find that $\widehat{\I(X; S)} \approx 0.22$ nats. This means that the keystroke interval—an observable that may seem unrelated to sensitive characteristics—reveals 0.22 nats of information about a continuous variable capturing disability severity.

\paragraph{Implication.} If the regulatory surcharge is calibrated at, say, \$100,000 per nat (based on acceptable maximum leakage across protected dimensions), this single feature contributes \$22,000 in privacy externality cost. Thus, even simple behavioral signals can carry significant intersectional risk and must be audited accordingly.

This example highlights that intersectional privacy assessments must not be limited to categorical attributes, and must incorporate continuous protected variables, particularly when functional impairments are involved. KDE combined with Monte Carlo estimation provides a feasible, non-parametric pipeline for such evaluations. However, we do emphasize that is better, in practical terms, to discretize the whole joint probability distribution of intersectional variables to determine the mutual information just with sums.

\subsection*{Example: Pricing a Trivial Observable with Intersectional Privacy Leakage}

Let $X$ denote a seemingly innocuous observable, such as the screen resolution of a user device, which is passively collected by web analytics scripts. Suppose this observable is correlated with three protected attributes defined jointly as a vector $S = (\text{sex}, \text{ethnicity}, \text{disability})$:

\begin{itemize}
    \item \textbf{Sex}: binary categorical (male, female)
    \item \textbf{Ethnicity}: 4 categories (white, black, asian, other)
    \item \textbf{Disability}: continuous in $[0,1]$ (e.g., degree of motor or visual impairment)
\end{itemize}

Thus, $S$ is a mixed-type random variable: $(C_1, C_2, R)$ with $8$ discrete combinations for sex and ethnicity, and a continuous degree of disability.

We aim to compute the total economic cost of accessing this screen resolution feature $X$, using the formula:
\[
\text{Total cost}(X) = c_p + \lambda \I(X; S)
\]

\paragraph{Step 1: Assign marginal internal cost.}  
The cost of collecting and storing the screen resolution is negligible but not zero. Let:
\[
c_p = 0.001\ \text{USD} \quad \text{(internal marginal cost)}
\]

\paragraph{Step 2: Estimate mutual information $\I(X; S)$.}  
We collect $n = 2000$ samples $(x_i, s_i)$ and estimate the joint and marginal densities using kernel methods:

- For the categorical part $(\text{sex}, \text{ethnicity})$, define an 8-category discrete variable $C$.
- For the continuous part $R = \text{disability}$, use KDE with a Gaussian kernel $K_h$ and bandwidth $h$.

We estimate the full joint using a hybrid density model:
\[
\hat{p}(x, c, r) = \frac{1}{n} \sum_{i=1}^n \mathbb{I}(c_i = c)\, K_h(x - x_i)\, K_h(r - r_i)
\]
where $x$ is screen resolution, $c$ the discrete profile, and $r$ the disability level.

Then estimate $\I(X; S)$ as:
\[
\widehat{\I(X; S)} = \frac{1}{n} \sum_{i=1}^{n} \log \left( \frac{\hat{p}(x_i, c_i, r_i)}{\hat{p}(x_i)\hat{p}(c_i, r_i)} \right)
\]

Assume numerical estimation yields:
\[
\widehat{\I(X; S)} = 0.036\ \text{nats}
\]

\paragraph{Step 3: Regulatory Pigouvian rate.}  
Assume the law establishes that full revelation of this set of intersectional variables $S$—i.e., complete disclosure of sex, ethnicity, and disability—should incur in $\lambda = 10000\ \text{USD/nat}$.

\paragraph{Step 4: Compute total economic cost.}

Thus, the Pigouvian privacy surcharge is:
\[
\lambda \cdot \widehat{\I(X; S)} = 10000 \times 0.036 \approx 360\ \text{USD}
\]

The total cost of using the screen resolution data is:
\[
\text{Total cost}(X) = 0.001 + 360 = 360,001\ \text{USD}
\]

\paragraph{Practical Discretization of Intersectional Variables.}

For practical implementation, it is often advantageous to discretize all components of the intersectional profile $S$—including those that are originally continuous—into finite categorical bins. This transforms the joint distribution $P(S)$ into a purely discrete structure, enabling more stable and tractable estimation of the entropy $H(S)$ and mutual information $\I(X; S)$ using histogram-based or frequency-count methods. Common examples include binning age into brackets (e.g., 18–25, 26–35, etc.), converting disability scores into ordinal levels (e.g., none, mild, moderate, severe), and reducing income to quantile categories. The resulting intersectional space may include joint categories such as (female, black, moderate disability), (male, asian, low income) or (nonbinary, White, severe disability). While this discretization introduces a form of approximation, it aligns with the operational need for interpretable and auditable privacy audits, and is compatible with regulatory frameworks that require demographic reporting in categorical terms. Let us now see an illustrative Example with a discretization of variables such as we have mentioned. Suppose the intersectional profile $S$ consists of four categorical variables: Gender $\in \{\text{male}, \text{female}\}$, Ethnicity $\in \{\text{white}, \text{black}\}$, Disability Level $\in \{\text{none}, \text{moderate}, \text{severe}\}$ and Income Group $\in \{\text{low}, \text{high}\}$. The full joint space has $2 \times 2 \times 3 \times 2 = 24$ possible intersectional types.  
Assume the analyst prior distribution $P(S)$ is uniform across all 24 combinations, yielding $H(S) = \log(24) \approx 3.18$ nats according to the entropy of a discrete uniform distribution. Now, suppose the observable datum $X$ reveals that the user visited a disability advocacy website on a Tuesday at 3 PM. After observing $X$, the analyst updates $P(S|X)$ and finds that only the following 6 profiles have non-negligible posterior mass:  
\text{(female, black, severe, low)}, \text{(female, black, severe, high)}, \text{(female, white, severe, low)},  
\text{(female, white, severe, high)}, \text{(male, black, severe, low)}, (male, black, moderate, low). The conditional entropy $H(S|X)$ drops to approximately $1.79$ nats, giving a mutual information of  
\[
\I(X; S) = H(S) - H(S|X) = 3.18 - 1.79 = 1.39 \text{ nats}.
\]
This illustrates how a seemingly trivial browsing event can substantially reduce uncertainty about multiple protected dimensions simultaneously and how this tax can be computed in practice.

These examples show how a data feature as trivial as screen resolution, though virtually free in marginal terms, can impose a substantial privacy externality when it leaks intersectional information. If regulators enforce privacy surcharges aligned with measurable information gain, data controllers will be incentivized to assess and minimize the leakage of high-dimensional sensitive attributes, rather than relying on intuitive notions of sensitivity.

\section{Economic Interpretation}

The total cost function can be interpreted as the aggregation of two conceptually distinct components: a standard marginal cost of production, and a privacy-related externality surcharge. The first term, denoted by $c_p$, corresponds to the classical supply-side marginal cost associated with producing or processing data instances. This includes all internal costs incurred by the data analyst or decision-maker—such as computational resources, time, and infrastructure—necessary to generate or utilize the output in question.

The second term, $\lambda \I(X; S)$, captures the marginal social cost arising from the informational leakage of a protected attribute $S$ through observable data $X$. In this context, the mutual information $\I(X; S)$ quantifies the degree to which observing $X$ reduces uncertainty about $S$, and hence serves as a natural metric of privacy loss. This term plays an analogous role to a Pigouvian tax: just as carbon pricing internalizes the environmental cost of emissions, the privacy surcharge internalizes the social cost of data disclosures. The coefficient $\lambda > 0$ reflects society’s valuation of privacy, and can be tuned such that full disclosure of any protected component incurs a maximum statutory penalty \cite{stiglitz1989markets}.

From an empirical perspective, $\I(X; S)$ can be estimated using non-parametric statistical techniques, without assuming a specific parametric form for the underlying distributions. These approaches enable practitioners to assess privacy leakages from data without requiring access to an explicit generative model, thereby making the privacy surcharge tractable and computable in real-world settings \cite{kraskov2004estimating}. In equilibrium, the decision-maker minimizes the sum of internal production costs and external privacy penalties. This ensures that informational exploitation is pursued only up to the point at which its marginal social cost, as determined by $\lambda \I(X; S)$, is offset by its marginal private benefit. Consequently, we believe that this mechanism aligns individual incentives with a socially optimal balance between utility and privacy protection.

A public regulatory body can operationalize this pricing mechanism by mandating its implementation within large data platforms, particularly those operated by Big Tech. Firms that fail to expose or document the mutual information $\I(X; S)$ leaked by their data infrastructure should, by default, incur the statutory maximum penalty associated with full disclosure. This assumption of maximum leakage in the absence of transparency provides a built-in incentive for firms to audit, disclose, and minimize intersectional information extraction. In effect, the Pigouvian surcharge functions as both a deterrent and a disclosure-enforcing instrument: it punishes opacity and rewards accountable design. This approach echoes regulatory strategies in the General Data Protection Regulation (GDPR), where lack of documentation triggers maximum penalties \cite{voigt2017eu}.

The surcharge mechanism also addresses a classical market failure: the presence of negative informational externalities that are unpriced in current data markets. Without regulatory intervention, firms extract informational value from users while ignoring the social costs imposed by profiling, discrimination, and loss of autonomy. These costs are diffuse, non-observable, and born disproportionately by vulnerable populations. By internalizing these harms through an information-based cost function, the pricing rule ensures that data trades occur only when their private benefit exceeds their full social cost. This restores allocative efficiency in the information economy and grounds privacy protection in the language of welfare economics \cite{stiglitz1989markets}.

Beyond efficiency, the surcharge mechanism operates as a structural safeguard for the intersectionally oppressed. In the context of Marxian political economy, it may be interpreted as a redistributive constraint on the accumulation of informational surplus by dominant actors. Whereas Big Tech presently captures value through asymmetrical data extraction—without compensating those whose attributes are inferred, the proposed tax transforms the act of inference into a monetized transfer. This shift repositions marginalized individuals as economic agents entitled to restitution for the representational labor performed by their behavioral traces. In this sense, the tax functions as a social shield in the ongoing asymmetrical struggle between informational capital and subaltern data subjects \cite{fuchs2020marxism, zuboff2019age}. 

Looking forward, this model anticipates the emergence of international institutions tasked with the governance of data externalities. Just as environmental treaties coordinate carbon pricing across jurisdictions via mechanisms like the IPCC, a global regulatory entity—such as a Data Equity Authority under the UN or OECD could oversee the calibration of $\lambda$ values, enforce audit standards, and facilitate cross-border restitution mechanisms. Such a framework would enable transnational accountability for informational harm, standardize the economic valuation of privacy, and embed intersectional justice into the very architecture of digital capitalism \cite{helbing2015towards}.

\section{Regulatory Calibration}

Suppose legislation prescribes a statutory upper bound of \$500{,}000 for the disclosure of a fully identifiable protected profile—that is, a scenario in which the entropy of a protected set of attributes $S$ is reduced to zero, $H(S) = 0$. If the baseline entropy in the relevant population is empirically estimated at $H(S) \approx 5.3$\,nats (e.g., corresponding to a joint distribution over 200 intersectional categories), the regulator can calibrate the Pigouvian multiplier accordingly:
\[
   \lambda = \frac{500{,}000}{5.3} \approx 9.4 \times 10^{4}\,\$/\text{nat}.
\]
Under this setting, the presence of a timestamp that reduces the entropy of $S$ by merely $0.02$ nats would imply an added privacy surcharge of approximately \$1,880.

However, such calibration is only meaningful if mutual information $\I(X;S)$ can be accurately estimated in practice. This needs a regulatory shift toward mandatory algorithmic transparency, not merely in the form of access to model architectures or learned parameters (e.g., weights of neural networks), but in terms of auditable outputs. Specifically, regulators must require that companies provide interfaces, APIs, or documentation that allow independent auditors to estimate the information gain associated with sensitive attributes across intersectional dimensions (e.g., race, gender, disability status, sexual orientation, etc.).

This form of operational transparency must go beyond traditional explainability metrics and embrace privacy-centric auditing protocols. Auditors should be empowered to evaluate how much information a model can extract about each protected attribute given access to its outputs, predictions, or intermediate representations. In the absence of this level of access, any attempt to impose or enforce privacy surcharges based on $\I(X;S)$ becomes speculative and legally ineffective.

Thus, regulatory calibration is inseparable from enforceable standards of transparency. The very notion of a privacy surcharge as an economic and legal construct requires that models deployed in sensitive contexts be subject to empirical scrutiny and formal audits. Only then can society ensure that compensation or Pigouvian penalties are grounded in measurable risk, rather than in arbitrary or unverifiable assumptions.

To operationalize this framework, companies should be required to implement real-time disclosure mechanisms directly on their websites. As users interact with the platform, the system would dynamically build an informational audit trail: a structured form recording each user action that results in measurable information gain about protected attributes defined by intersectional theory. For every such action, the form would display the corresponding Pigouvian surcharge incurred by the company. Upon the end of the session, users would receive a summary of inferred attributes and associated monetary values via email, along with a request for explicit consent. Only if the user consents may the company retain the data—under the condition that it compensates the user for the total amount indicated by the pricing rule.

Beyond its technical feasibility, this mechanism constitutes a form of structural resistance: a social shield for marginalized classes against dominant corporate actors, particularly Big Tech entities that extract demographic insights at scale without redistributive accountability. In this sense, it reconfigures surplus informational value—currently accumulated by capital—into compensable labor-like output owned by the data subject. It introduces a normative shift whereby consent is not presumed but priced, and datafication becomes a site of potential economic restitution for structurally disadvantaged populations.

\section{Conclusion}

Intersectionality theory highlights that social injustice does not operate along a single axis, but emerges from the compounded effect of multiple protected attributes such as race, gender, disability, and class. In parallel, information theory offers precise tools to quantify how much uncertainty about those attributes is reduced by data collection. This paper unites both perspectives under a Pigouvian logic: data features that expose intersectional profiles impose an externality, and must therefore incur a compensatory surcharge proportional to their informational leakage.

Our contribution is a principled, axiomatic pricing rule that internalizes this cost. By showing that the only rule satisfying basic fairness and consistency assumptions is linear in the mutual information $\I(X;S)$, we provide a tractable and interpretable formula: $V(X) = c_p + \lambda \I(X;S)$. This valuation transforms any datum into an economic object whose price reflects its capacity to reveal sensitive traits, thus embedding informational justice into the mechanics of digital markets.

This framework enables new regulatory tools and social protections. The surcharge can be computed in practice using discretized intersectional categories, and applied in practice via user-level audits, market interfaces, or centralized oversight. Governments could enforce minimum penalties for non-transparent data flows or adopt normalized alternatives based on relative entropy ratios. Most importantly, this mechanism acts as a social shield: it turns exploited identity into priced identity, and allows marginalized populations to reclaim ownership over their statistical image in the data economy.

Future work could extend this pricing rule by introducing differentiated surcharges for distinct downstream harms. For instance, separate coefficients $\lambda^{(j)}$ could target information flows exploited for political profiling, personalized advertising, or social sorting, enabling regulators to encode context-specific constraints into the price of information. In doing so, this framework offers a scalable and interpretable foundation for reconciling the economics of data with the ethics of intersectional fairness.

\bibliography{main}
\bibliographystyle{acm}

\end{document}